\documentclass{iopart}
\usepackage{iopams}
\usepackage[dvips]{color}
\usepackage{hyperref}

\begin{document}
\title[Gauge-gravity]
{A note on a gauge-gravity relation and functional determinants}

\author{R Aros, D E D\'{\i}az and A Montecinos}

\address{Universidad Andr\'es Bello,
Departamento de Ciencias F\'{\i}sicas,
Rep\'ublica 220, Santiago, Chile}
\ead{raros,danilodiaz,alejandramontecinos@unab.cl}

\begin{abstract}

We present a refinement of a recently found gauge-gravity relation between one-loop effective actions: on the gauge side, for a massive charged scalar in 2d dimensions
in a constant maximally symmetric electromagnetic field; on the gravity side, for a massive spinor in d-dimensional (Euclidean) anti-de Sitter space.
The inclusion of the dimensionally regularized volume of AdS leads to complete mapping within dimensional regularization. In even-dimensional AdS, we get a small correction to the original proposal; whereas in odd-dimensional AdS, the mapping is totally new and subtle, with the `holographic trace anomaly' playing a crucial role.

\end{abstract}



\vspace{10mm}

Functional determinants of geometric differential operators, such as the Laplacian and Dirac's operator, are ubiquitous
in mathematical and theoretical physics (cf.~\cite{Branson:1993xx,Dunne:2007rt}). In particular, in quantum field theory, the one-loop effective action is given by the determinant of the `propagating' differential operator. Exact results for these determinants in gauge theory date back at least to the works of Heisenberg and Euler and of Weisskopf, for constant electromagnetic backgrounds. In special gravitational backgrounds, such as AdS, the high degree of symmetry also allows for exact expressions.

Even though such one-loop effective actions have been known for quite some time, it was only very recently that a remarkable gauge-gravity relation was identified~\cite{Basar:2009rp}. On the gauge side, one considers the one-loop effective action for a massive charged scalar in 2d dimensions in a constant maximally symmetric electromagnetic field; on the gravity side, the one-loop effective action for a massive spinor in d-dimensional (Euclidean) anti-de Sitter space.
The observation was inspired by factorization of graviton amplitudes in terms of gauge amplitudes and the prospects for inclusion of one-loop effects, while the similarity was revealed by the presence of Barnes' multiple gamma function~\cite{Ruijsenaars,FriedmanRuijsenaars}.
Several possible directions in which the observation might be extended include relaxing the maximal symmetry of the electromagnetic field or its homogeneity, and considering higher-loops. In addition, one might also contemplate including finite temperature effects.

Our concern here, however, is with a seemingly overlooked detail, namely, that the mapping is only established for even-dimensional AdS. On the gauge side, as long as the number of dimensions is twice larger that of AdS, the effective action will be given in terms of Barnes' multiple gamma function. In odd dimensional AdS, in turn, the renormalized effective Lagrangian in polynomial in the mass parameter and little hope is left to meet Barnes' multiple gamma in this case.
In this communication we show how to amend this apparent mismatch by taking into account the dimensionally regularized volume of AdS space.

\section*{Gauge side}

On the gauge side, one considers a massive ($M$) charged scalar field $\phi$ in flat Euclidean $2d-$dimensional space in the constant background of a maximally symmetric electromagnetic field\footnote{We refer to~\cite{Basar:2009rp} for conventions and references.}. The field strength can be written in block form
\begin{equation}
F_{\mu\nu}=f\,\mbox{diag}_d\left\{\left(\begin{array}{c c} 0&1 \\ -1 & 0 \end{array}\right)\right\}~.
\end{equation}
The \textit{dramatis persona} on the gauge side is the functional determinant arising in the one-loop effective action
\begin{equation}
S^{2d}_{gauge}=\ln \det\{-D^2+M^2\}~.
\end{equation}
Standard manipulations lead to an expression in terms of the Green's function $G$, schematically $(-D^2+M^2)\,G=-1$,
\begin{equation}
S^{2d}_{gauge}=-\int^{M^2}\,\tr G^{2d}~,
\end{equation}
where trace $\tr$ means volume integral (large box of length $L$) of the diagonal entry $G^{2d}(x,x)$.
The heat kernel gives a useful representation of this diagonal entry, and a starting point to deal with the inherent short-distance divergence,
\begin{equation}
-\int^{M^2}\tr G^{2d}=L^{2d}\int^{\infty}_0\;\frac{ds}{s}\frac{1}{(4\pi s)^d}\;\left(\frac{fs}{\mbox{sh}(fs)}\right)^d\,e^{-sM^2}~.
\end{equation}
The divergence at the lower integration value of the proper-time integral could be cured if the dimension $d$ were negative enough. This observation is the key to dimensional regularization where the dimension is now $D=d-\epsilon$  and one can work out the value of the effective action in closed form
\begin{equation}\label{gauge-eff}
S^{2D}_{gauge}=-(\frac{fL^2}{2\pi})^D\;\Gamma(1-D)\,\int^{\frac{M^2}{2f}}\,d\mu\;
\frac{\Gamma(D/2+\mu)}{\Gamma(1-D/2+\mu)}~.
\end{equation}
The divergence as $D\rightarrow d$, \textit{i.e.} $\epsilon\rightarrow 0$, comes entirely from the $\Gamma(1-D)$ factor for any integer value of $d$, even or odd.

Now, the main observation in~\cite{Basar:2009rp} is related to the presence of Barnes' multiple gamma function in the renormalized value of this expression. A quick way to get in touch with Barnes' multiple gamma function is to go back to the heat kernel representation and perform a `Weierstrass regularization', that is, subtract successive terms of the Taylor expansion of the integrand to achieve convergence in the lower integration limit.
We stick to the conventions in~\cite{Ruijsenaars, FriedmanRuijsenaars} for Barnes' multiple gamma function~(\ref{Barnes}), modulo renormalization scheme dependent terms which are polynomial and logarithmic in $m$, and obtain the finite renormalized value
\begin{equation}\label{gauge-result}
S^{2d}_{gauge}=(\frac{fL^2}{2\pi})^d\cdot\ln\Gamma_d(\frac{d}{2}+\frac{M^2}{2f})~.
\end{equation}

Alternatively, expanding in $\epsilon$, $D=d-\epsilon$, and after canceling the pole from $\Gamma(1-D)$ with the linear-in-$\epsilon$ term in the expansion of quotient of gamma functions involving the mass, one gets

\begin{eqnarray}
\fl S^{2d}_{gauge}&=(\frac{fL^2}{2\pi})^d\;\frac{(-1)^{d-1}}{2(d-1)!}\int^{\frac{M^2}{2f}}\,d\mu\;\frac{\Gamma(\frac{d}{2}+\mu)}{\Gamma(1-\frac{d}{2}+\mu)}
\,\left(\psi(\frac{d}{2}+\mu)+\psi(1-\frac{d}{2}+\mu)\right)\,+...\nonumber\\
\fl &=(\frac{fL^2}{2\pi})^d\;\frac{(-1)^{d-1}}{(d-1)!}\int^{\frac{M^2}{2f}}\,d\mu\;\left\{\prod_{j=0}^{d-2}
(\frac{d}{2}+\mu-1-j)\right\}\psi(\frac{d}{2}+\mu)\,+...\nonumber\\
\fl &=(\frac{fL^2}{2\pi})^d\cdot\ln\Gamma_d(\frac{d}{2}+\frac{M^2}{2f})+...
\end{eqnarray}
Any subtraction or renormalization prescription will eventually lead to Barnes' multiple gamma, modulo the aforementioned logarithmic and polynomial terms related to a
renormalization scheme ambiguity.
\section*{Gravity side}

Let us turn now to the gravity side to consider a massive ($m$) Dirac spinor minimally coupled to gravity in the Euclidean $AdS_d$ background.
The \textit{dramatis persona} on the gravity side is the functional determinant in the one-loop effective action
\begin{equation}
S^{d}_{gravity}=-\ln \det\{\slash{\!\!\!\nabla}+m\}~.
\end{equation}
Again, standard manipulations bring in the Green's function, $(\slash{\!\!\!\nabla}+m)\mathcal{D}=-\mathbb{I}$,
\begin{equation}
S^{d}_{gravity}= \int^m \tr \mathcal{D}^d~,
\end{equation}
where now also the spinor indices are traced out.

The spinor propagator in $AdS_d$ is a well known quantity~\cite{Camporesi:1990xx}, involving Gauss' hypergeometric, and its diagonal entry can be worked out in dimensional regularization
\begin{equation}
\mathcal{D}^D(x,x)=-\,\frac{\mathbb{I}}{\sqrt{\Lambda}}\,(\frac{\Lambda}{4\pi})^{D/2}\;\Gamma(1-\frac{D}{2})\;
\frac{\Gamma(\frac{D}{2}+\frac{m}{\sqrt{\Lambda}})}{\Gamma(1-\frac{D}{2}+\frac{m}{\sqrt{\Lambda}})}~,
\end{equation}
with $\mathbb{I}$ denoting the identity matrix in spinor indices, whose dimensionality is $2^{[D/2]}$.
The analogy with the gauge result is already apparent, but there is a seemingly small mismatch in the gamma factor: $\Gamma(1-\frac{D}{2})$ instead of $\Gamma(1-D)$. This makes a tiny difference when $d=even$, both expressions have a pole and the finite piece involves Barnes' multiple gamma.
However, when $d=odd$, there is a huge difference and there is no mapping of gauge and gravity quantities. On the gauge side there is still a pole, but the gravity side is finite in the limit $D\rightarrow d$.
Our contribution in this note is to amend this discrepancy by taking into account the dimensional regularized volume of Euclidean AdS which is finite for $d=even$, whereas for $d=odd$ one finds the missing pole in the above discussion.

To compute the dimensional regularized volume of $\mathbb{H}^D$, start with the metric
\begin{equation}
ds^2= \frac{R^2}{r^2}\left[dr^2+(1-r^2)^2\,g_o\right]~,
\end{equation}
where $4\,g_o$ is the round metric on the sphere $S^{D-1}$ and $R=1/\sqrt{\Lambda}$ is the radius of AdS.
The volume is then~\cite{Diaz:2007an}
\begin{eqnarray}
vol(\mathbb{H}^D)&=R^D\,2^{1-D}\,vol(S^{D-1})\,\int^1_0 dr\, r^{-D}\,(1-r^2)^{D-1}\nonumber\\
&=
\,\pi^{\frac{D}{2}-\frac{1}{2}}\,R^D\,\Gamma(\frac{1}{2}-\frac{D}{2})~.
\end{eqnarray}

Notice that the volume has a pole for $D\rightarrow odd$ and is finite for $D\rightarrow even$, so that when combined with the effective Lagrangian there is always a divergence as $D$ goes to the physical dimension, even or odd, as in the gauge side.
Using Legedre's duplication formula for the product of the gamma's we end up with
\begin{equation}
S^{D}_{gravity}=-2^{[D/2]}\,\Gamma(1-D)\,\int^{\frac{m}{\sqrt{\Lambda}}}\,d\mu\;
\frac{\Gamma(\frac{D}{2}+\mu)}{\Gamma(1-\frac{D}{2}+\mu)}~.
\end{equation}

Now one can appreciate the mapping to the gauge one-loop effective action~(\ref{gauge-eff}).
\subsection*{Even dimensions: $d=2n$}

Now we go to the physical dimension $d=2n$. The volume asymptotics, $D=2n-\epsilon$ as $\epsilon\rightarrow 0$, is finite
\begin{equation}
vol(\mathbb{H}^D)=\mathcal{V}_{2n} + o(1)~,
\end{equation}
with $\mathcal{V}_{2n}=\pi^{n-1/2}\,R^{2n}\,\Gamma(\frac{1}{2}-n)=(-1)^n\,\frac{\pi^{n+1/2}}{\Gamma(n+\frac{1}{2})}\,R^{2n}$.

Notice that our renormalized volume differs by a factor $(-1)^n$ from the volume element used in~\cite{Basar:2009rp}. We welcome this extra factor for it makes more precise the gauge-gravity mapping.

The important cancelation occurs in the expansion of the effective Lagrangian, very much as in the gauge side,
\begin{equation}\fl S^{2n}_{gravity}=\mathcal{V}_{2n}\,(\frac{\Lambda}{4\pi})^n\;\frac{(-2)^{n+1}}{(n-1)!}\int^{\frac{m}{\sqrt{\Lambda}}}\,d\mu\;\left\{\prod_{j=0}^{2n-2}
(n+\mu-1-j)\right\}\psi(n+\mu)\,+...
\end{equation}

After including the volume factor and using the duplication formula for the gamma, we finally get
\begin{equation}
S^{2n}_{gravity}= 2^n\cdot\ln\Gamma_{2n}(n+\frac{m}{\sqrt{\Lambda}})~,
\end{equation}
to be compared with the gauge result~(\ref{gauge-result}).

\subsection*{Odd dimensions: $d=2n+1$}

This case is qualitatively different from the above; now the volume gets mixed with the subleading term in the $\epsilon -$expansion of the effective Lagrangian.
The volume asymptotics $D=2n+1-\epsilon$ as $\epsilon\rightarrow 0$ goes as
\begin{equation}
vol(\mathbb{H}^D)=\frac{1}{\epsilon}\,\mathcal{L}_{2n+1}+ \mathcal{V}_{2n+1} + o(1)~.
\end{equation}
The precise value of the renormalized volume $\mathcal{V}_{2n+1}$ will not be relevant, the important terms is the `integrated holographic trace anomaly'~\cite{HS98,Gra99,Albin:2005} which  is unambiguously given by $\mathcal{L}_{2n+1}=(-1)^n\frac{2\pi^n}{\Gamma(n+1)}\,R^{2n+1}$.

The effective Lagrangian is now finite as $D\rightarrow 2n+1$, and polynomial in $m$. The nontrivial part, which eventually leads to Barnes' multiple gamma function, comes from the cancelation of the pole in the volume against the linear-in-$\epsilon$ term in the expansion of the effective Lagrangian\footnote{An analogous `conspiracy' was also crucial in establishing a holographic formula for functional determinants AdS~\cite{Diaz:2007an}.}
\begin{eqnarray}\fl S^{2n+1}_{gravity}&=\mathcal{L}_{2n+1}\,2^n\,(\frac{\Lambda}{4\pi})^{n+\frac{1}{2}}\,\Gamma(\frac{1}{2}-n)\times\nonumber\\
\fl &\times\int^{\frac{m}{\sqrt{\Lambda}}}\,d\mu\;
\frac{\Gamma(n+\frac{1}{2}+\mu)}{\Gamma(\frac{1}{2}-n+\mu)}
\,\left(\psi(n+\frac{1}{2}+\mu)+\psi(\frac{1}{2}-n+\mu)\right)\,+...\\
\fl &=\mathcal{L}_{2n+1}\,\frac{(-2)^{-n}\pi^{\frac{1}{2}-n}\Lambda^{n+\frac{1}{2}}}{\Gamma(n+\frac{1}{2})}
\int^{\frac{m}{\sqrt{\Lambda}}}\,d\mu\;
\left\{\prod_{j=0}^{2n-1}
(n-\frac{1}{2}+\mu-j)\right\}\psi(n+\frac{1}{2}+\mu)+...\nonumber
\end{eqnarray}
A little algebraic manipulations, and the final answer reads
\begin{equation}
S^{2n+1}_{gravity}= 2^n\cdot\ln\Gamma_{2n+1}(n+\frac{1}{2}+\frac{m}{\sqrt{\Lambda}})~.
\end{equation}

In all, on the gravity side the one-loop effective action after inclusion of the volume term in $d$ dimensions, even and odd, reads
\begin{equation}
S^d_{gravity}= 2^{[d/2]}\cdot\ln\Gamma_d(\frac{d}{2}+\frac{m}{\sqrt{\Lambda}})
\end{equation}
in remarkable correspondence with the gauge result~(\ref{gauge-result}) and with $d=odd$ and $d=even$ on equal footing.

\ack We kindly acknowledge E. Friedman and S. Ruijsenaars for a useful conversation on Barnes' zeta and
gamma functions. D.E.D. wishes to thank H. Dorn for valuable discussions.
This work was partially funded through Fondecyt-Chile 3090012, UNAB AR-01-09 and UNAB DI 2009/04.
\appendix

\section{Useful identities}

\begin{itemize}
\item Integral representation of Barnes' multiple gamma, which effectively performs the Weierstrass regularization,
\begin{eqnarray}\label{Barnes}
\fl \ln\Gamma_N(w)&=\int^{\infty}_0 \frac{dt}{t}\left( e^{-wt}\prod^N_{j=1}\frac{1}{1-e^{-a_jt}}\right.\nonumber\\
\fl &\left.-t^N \sum^{N-1}_{k=0}
\frac{(-t)^k}{k!}B_{N,k}(w)- \frac{(-1)^N}{N!}e^{-t}B_{N,N}(w)\right)~,
\end{eqnarray}
with $B_{N,k}(w)$ being Bernoulli-type polynomials~\cite{Ruijsenaars,FriedmanRuijsenaars}.
\vspace{4mm}
\item Gamma function, duplication formula
\begin{equation}
\Gamma(z)\,\Gamma(z+\frac{1}{2})=2^{1-2z}\,\Gamma(\frac{1}{2})\,\Gamma(2z)
\end{equation}
\vspace{4mm}
\item Gamma function, reflection formula
\begin{equation}
\Gamma(\frac{1}{2}+z)\,\Gamma(\frac{1}{2}-z)=\frac{\pi}{\cos \pi z}
\end{equation}
\end{itemize}

\vspace{0.5cm}

\section*{References}

\providecommand{\href}[2]{#2}\begingroup\raggedright\endgroup

\begin{thebibliography}{10}

\bibitem{Branson:1993xx}
 T.~Branson,
 {\it The functional determinant,}
 {\em Lecture Note Series, Vol. 4,} Global Analysis Research Center,
 Seoul National University, 1993.
\bibitem{Dunne:2007rt}
  G.~V.~Dunne,
  {\it Functional Determinants in Quantum Field Theory,}
  {\em J. Phys. A}{\bf 41} (2008) 304006
  [arXiv:0711.1178 [hep-th]].
\bibitem{Basar:2009rp}
  G.~Basar and G.~V.~Dunne,
  {\it A Gauge-Gravity Relation in the One-loop Effective Action,}
  {\em J. Phys. A}{\bf 43} (2010) 072002,
  [\href{http://xxx.lanl.gov/abs/0912.1260}{{\tt hep-th/0912.1260}}].
\bibitem{Camporesi:1990xx}
R.~Camporesi,
 {\it The Spinor heat kernel in maximally symmetric spaces,}
 {\em Commun. Math. Phys.} {\bf 148} (1992) 283.
\bibitem{Diaz:2007an}
D.~E.~Diaz and H.~Dorn, {\it {Partition functions and double-trace deformations
  in AdS/CFT}},  {\em JHEP} {\bf 05} (2007) 046,
  [\href{http://xxx.lanl.gov/abs/hep-th/0702163}{{\tt hep-th/0702163}}].
\bibitem{Ruijsenaars}
S.~N.~M.~Ruijsenaars, {\it On Barnes' multiple zeta and gamma functions},
{\em Advances in Mathematics} {\bf 156} (2000) 107--132.
\bibitem{FriedmanRuijsenaars}
E.~Friedman and S.~N.~M.~Ruijsenaars, {\it Shintani-Barnes zeta and gamma functions},
{\em Advances in Mathematics} {\bf 187} (2004) 362--395.
\bibitem{HS98}
M.~Henningson and K.~Skenderis, ``The holographic Weyl anomaly,''
JHEP {\bf 9807} (1998) 023 [arXiv:hep-th/9806087];
``Holography and the Weyl anomaly,'' Fortsch.\ Phys.\  {\bf 48}
(2000) 125 [arXiv:hep-th/9812032].
\bibitem{Gra99}
C.~R.~Graham, ``Volume and area renormalizations for conformally
compact Einstein metrics,'' Rend.\ Circ.\ Mat.\ Palermo (2) Suppl.
No. 63 (2000) 31 [arXiv:math.DG/9909042].
\bibitem{Albin:2005}
  P.~Albin, {\it {Renormalizing Curvature Integrals on Poincare-Einstein Manifolds}},
  \href{http://xxx.lanl.gov/abs/math.DG/0504161}{{\tt math.DG/0504161}}.
\end{thebibliography}
\end{document}